\documentclass[a4paper,fleqn,final]{cas-sc}
\usepackage{mathtools} 
\usepackage{mathrsfs}
\usepackage{color}
\usepackage{fancyhdr}
\usepackage{bbold}
\usepackage[numbers, sort&compress]{natbib}
\usepackage{diagbox} 
\usepackage[T1]{fontenc}
\usepackage{caption}
\usepackage{hyperref}
\usepackage{bm}
\usepackage{comment}
\usepackage{orcidlink} 

\definecolor{dark green}{rgb}{0.00, 0.39, 0.00}

\newcommand\aap{A\&A}                
\newcommand\apj{ApJ}                 
\newcommand\araa{ARA\&A}             
\newcommand\pasj{PASJ}               
\newcommand\jcap{J.~Cosmology Astropart. Phys.} 
\newcommand\mnras{MNRAS}             
\newcommand\prd{Phys. Rev.~D}        
\newcommand\prl{Phys. Rev.~Lett.}    
\providecommand{\apss}{Ap\&SS}
\providecommand{\aap}{A\&A}
\providecommand{\mnras}{MNRAS}
\providecommand{\prd}{Phys. Rev. D}
\providecommand{\prl}{Phys. Rev. Lett.}
\providecommand{\jcap}{JCAP}
\providecommand{\araa}{ARA\&A}
\providecommand{\apj}{ApJ}

\newcommand{\fref}[1]{Fig.~\ref{#1}}

\newcommand{\sref}[1]{Sec.~\ref{#1}}
\newcommand{\tref}[1]{Table~\ref{#1}}

\newcommand{\newc}{\newcommand}
\newcommand\be{\begin{equation}}
\newcommand\ee{\end{equation}}
\newcommand\ba{\begin{eqnarray}}
\newcommand\ea{\end{eqnarray}}

\newc{\tp}{\dot{\phi}}

\renewcommand{\(}{\left(}
\renewcommand{\)}{\right)}

\newcommand{\crit}{{\rm crit}}

\begin{document}
\let\WriteBookmarks\relax
\def\floatpagepagefraction{1}
\def\textpagefraction{.001}
\shorttitle{Cluster number counts in dark energy model with energy and momentum coupling to dark matter}
\shortauthors{\relax}
\title[mode = title]{Cluster number counts in dark energy model with energy and momentum coupling to dark matter}
                
\author[1, 3]{Chattree Wongsangwal}[orcid=0009-0004-8449-872X]
\ead{chattreew63@nu.ac.th}

\author[1]{Khamphee Karwan}[orcid=0009-0004-4993-4733]
\ead{khampheek@nu.ac.th}

\author[2]{Stharporn Sapa}[orcid=0009-0003-1495-4913]
\ead{stharporn@northern.ac.th}

\author[1]{Teeraparb Chantavat}[orcid=0000-0002-0259-1591]
\ead{teeraparbc@nu.ac.th}

\affiliation[1]{organization={The Institute for Fundamental Study, Naresuan University},
                city={Phitsanulok},
                postcode={65000}, 
                country={Thailand}}
\affiliation[2]{organization={Northern College},
                city={Tak},
                postcode={63000},
                country={Thailand}}
\affiliation[3]{organization={National Astronomical Research Institute of Thailand},
                city={Chiang Mai},
                postcode={50180}, 
                country={Thailand}}
\begin{abstract}
The impact of momentum transfer in coupled dark energy models on galaxy cluster number counts is investigated.
The extrapolated linear density contrast at collapse, computed using the spherical collapse model, is found to suppress as the strength of momentum transfer increases.
The cluster number counts are then calculated using available analytical halo mass functions.
For each redshift bin, the minimum halo mass limit used in the mass integration is determined by matching the predicted number counts from the $\Lambda$CDM model with results from eROSITA survey.
We find that the number of clusters is maximal in higher redshift bins, and that the number of clusters in the highest redshift bin is enhanced as the strengths of momentum and energy transfer increase.
The cluster number counts are enhanced due to momentum and energy transfer, even though momentum transfer suppresses the linear growth rate of matter perturbations on small scales, while energy transfer enhances it.
In general, it may be concluded that the enhancement of the cluster number counts is a consequence of the suppression of the extrapolated linear density contrast.
The model parameters are set according to the constraints in \cite{bestfit}.
The predicted number counts from the coupled dark energy model are much larger than the Poisson error of the eROSITA survey results.
This can be a consequence of uncertainties in computing the number counts for the coupled dark energy model using the spherical collapse model and the analytical mass functions.
\end{abstract} 	


\maketitle
\section{Introduction} \label{sec1}
Cosmic acceleration at late times, which is strongly supported by numerous observations, 
suggests the existence of an unknown form of dark energy \cite{Shinji:DE, Bamba_LCDM}.
A possible candidate for dark energy is the cosmological constant $\Lambda$ \cite{Carroll2}.
Even though the $\Lambda$CDM model fits current observational data well, 
it encounters a problem due to the extremely large discrepancy between the value of the cosmological constant predicted theoretically and that inferred from observations \cite{Sahni2, Weinberg}.
To alleviate this problem, dark energy is often assumed to be a dynamical energy component that can take the form of a scalar field dubbed quintessence \cite{quintessence1, quintessence2, quintessence5}.

Since the dynamics of quintessence are different from and independent of those of radiation and matter, the coincidence problem arises. 
This problem asks why the energy densities of dark energy and dark matter are of the same order at present \cite{quintessence3, coincidence, coincidence3}.

Interaction between dark energy and dark matter is a possible mechanism for alleviating the coincidence problem, 
where such coupling is expected to arise in the context of particle physics \cite{Wetterich, coupling}.
Coupling between the dark sectors can also arise from frame transformations of the gravitational action.
A conformal transformation leads to an interaction term $\rho_m \dot\phi$ in the evolution equation for the background quintessence field \cite{Amendola:1999qq, Thipaksorn:2022yul}, 
where $\rho_m$ is the matter energy density and $\dot\phi$ is the time derivative of the quintessence field $\phi$. 
This form of coupling generates energy transfer between the dark sectors and has been extensively investigated \cite{Wetterich, Amendola:1999qq, Pettorino2013, Ade:2015rim, operatorlambda, 2020PhRvD.101f3502D}.
However, it has several drawbacks.
Although a scaling solution corresponding to a matter-dominated epoch exists, it is followed by an accelerated epoch that does not agree with observational data; specifically, 
the total equation-of-state parameter is larger than the observed value \cite{Amendola:2006qi}.
Moreover, this type of coupling enhances the linear growth rate of CDM perturbations on small scales \cite{operatorlambda}, which exacerbates the $\sigma_8$ tension.

The $\sigma_8$ tension refers to the discrepancy between the value of $\sigma_8$ inferred from Cosmic Microwave Background (CMB) observations and that obtained from weak lensing analyses \cite{s8tension1, s8tension2}.
The $\sigma_8$ parameter is the present-day root-mean-square (RMS) amplitude of the linear matter density fluctuations within a comoving sphere of radius $8h^{-1}{\rm Mpc}$.
In addition, the $\Lambda$CDM model also suffers from the $H_0$ tension between CMB measurements \cite{Planck2015, 2023Univ....9..393V} and local determinations of the expansion rate of the Universe \cite{Riess:2018}.

Coupling between the dark sectors can also arise from momentum transfer, described by an interaction between the field derivative $\partial_\mu \phi$ and the CDM four-velocity $u^\mu$ in the action \cite{Pourtsidou:2013nha, Boehmer:2015kta, scaling}.
This momentum coupling can yield scaling solutions that describe an accelerated epoch following the matter-dominated epoch, consistent with observational data \cite{scaling}.
It has been shown that momentum coupling suppresses the growth rate of CDM perturbations on small scales, which may help alleviate the $\sigma_8$ tension \cite{Linton_2022}.
Observational constraints on coupled dark energy models with both energy and momentum transfer suggest that parameter regions corresponding to zero momentum transfer are excluded at the $2\sigma$ level, and that the mean value of $\sigma_8$ is reduced compared to $\Lambda$CDM \cite{bestfit}.

Coupling between dark energy and dark matter can leave imprints on large-scale structure, allowing observational data from galaxy surveys to constrain such interactions.
Recently, results from the Spectrum-Roentgen-Gamma/extended ROentgen Survey with an Imaging Array (SRG/eROSITA) all-sky survey have been released, including the redshift distribution of confirmed clusters \cite{2402.08458}.
Cluster number counts derived from catalogs in the Western Galactic Hemisphere, combined with other observational data, have been used to constrain the growth of structure \cite{2410.09499}.

In a simple approach, cluster number counts for a given cosmological model can be estimated by combining results from the spherical collapse model with the Press–Schechter or Sheth–Tormen mass functions (see, e.g., \cite{numbercount1, spcollapse8}).
Cluster number counts can be used to constrain cosmological parameters \cite{numbercount5, numbercount6, numbercount7} and the power spectrum \cite{numbercount3, numbercount4}.
The influence of coupling between dark energy and dark matter on cluster number counts has been investigated in \cite{numbercount1, numbercount2, spcollapse5}.
The number density of galaxy clusters for various dark energy potentials has been studied in \cite{count4}.

In this work, we investigate the impact of momentum coupling between the dark sectors on cluster number counts.

In \sref{sec2}, we present the coupled dark energy model with energy and momentum transfer considered in this work.
Spherical collapse and cluster number counts are discussed in \sref{sec3} and \sref{sec4}. We provide our concluding remarks in \sref{sec5}.

\section{Coupled Dark Energy with Energy and Momentum Transfer}
\label{sec2}
In this section, we review the essentials of the coupled dark energy model with energy and momentum transfer, as well as results from previous analyses in the literature.

\subsection{Coupled Dark Energy Model}
\label{sec2.1}

The scalar field dark energy with energy and momentum coupling to cold dark matter (CDM) can be described by the action \cite{scaling}
\be
{\cal S} 
= 
\int {d}^4 x \, \sqrt{-g}\,
\frac{M_{\rm pl}^2}{2}\,R
-\int {d}^{4}x \left[\sqrt{-g}\,\rho_m(n_m) + J_m^{\mu}\, \partial_{\mu} \ell \right]
+\int {d}^4 x\, \sqrt{-g}\,L(n_c,\phi,X,Z)\,,
\label{action}
\ee
where $M_{\rm pl} \equiv 1 / \sqrt{8\pi\,G}$, and $R$ is the Ricci scalar.
The second term in the Eq.~\eqref{action} is the Schutz-Sorkin action describing the matter components in the form of a barotropic perfect fluid,
whose dynamics are quantified by the energy density $\rho_m$ and the particle number current $J_m^{\mu}$.
The matter energy density depends on the number density $n_m$,
while the current is defined as
\be
J_m^\mu =  u_m^{\mu}\, n_m \sqrt{-g}.
\ee
The subscript ${}_m$ denotes matter components in the universe, including radiation, baryon, and CDM.
The $u_m^\mu$ is the four-velocity of the fluid.
In the Schutz-Sorkin action~\cite{Schutz-Sorkin}, 
$\ell$ is a Lagrange multiplier or Clebsch potential which is introduced to ensure the conservation of particle number.
The action for a scalar field dark energy is the last term of Eq.~\eqref{action}.
The kinetic term of the scalar field is expressed as $X \equiv -g^{\mu \nu}\, \nabla_{\mu}\phi\, \nabla_{\nu} \phi/2$. 
The momentum coupling between dark energy and dark matter is encoded in $Z \equiv u_c^{\mu} \,\nabla_{\mu} \phi$,
where $u_c^{\mu}$ is the four velocity of CDM.
The energy transfer between dark energy and dark matter arises due to the dependence of the Lagrangian for a scalar field on a number density $n_c$ of dark matter.
The Lagrangian for dark energy with the momentum and the energy coupling to dark matter can be expressed as~\cite{scaling}
\be
L(n_c,\phi,X,Z) = -f_1(\phi,X,Z)\, \rho_c (n_c) + f_2(\phi,X,Z)\,,
\label{Lin}
\ee
where $\rho_c$ is the energy density of CDM, $f_1$ and $f_2$ are arbitrary functions that describe the energy and momentum transfer, respectively.

Varying the action \eqref{action} with respect to the metric tensor, we obtain the Einstein field equation
\be
M_{\rm pl}^2 \,G_{\mu\nu}
=
T_{\mu\nu}^{(r)}
+ \left( 1+f_1 \right) T^{(c)}_{\mu\nu}
+ T_{\mu\nu}^{(b)} + T_{\mu\nu}^{(\phi)}\,,
\label{Ein}
\ee
where $G_{\mu\nu}$ is the Einstein tensor and superscripts ${}^{(r)}, {}^{(c)}, {}^{(b)}$ and ${}^{(\phi)}$ denote radiation, CDM, baryon, and scalar field, respectively.
The energy-momentum tensor for matter and scalar field are in the form of
\ba
T^{(m)}_{\mu\nu} 
&=& \left( \rho_m+P_m \right)\, u_{\mu}\, u_{\nu} + P_m\, g_{\mu \nu}\,,
\label{Tmmun}\\
T^{(\phi)}_{\mu \nu} 
&=& f_2\, g_{\mu \nu} - \rho_c \left( f_{1,X}\, \nabla_{\mu}\phi\, \nabla_{\nu}\phi + f_{1,Z}\,Z\, u_{\mu} u_{\nu} \right) + f_{2,X}\,\nabla_{\mu}\phi\, \nabla_{\nu}\phi + f_{2,Z}\, Z\, u_{\mu} u_{\nu}\,,
\label{Tmm-phi}
\ea
where $P_m$ is the pressure of the matter.
The energy-momentum tensor in Eq.~\eqref{Tmmun} satisfies the conservation equation~\cite{scaling}
\be
\nabla^{\mu} T_{\mu \nu}^{(m)} = 0\,,
\label{Tcon}
\ee
while the conservation equations for the scalar field and CDM satisfy
\be
\left( 1+f_1 \right)\, \nabla^{\mu} T_{\mu \nu}^{(c)}
+ T_{\mu \nu}^{(c)}\, \nabla^{\mu} f_1
+ \nabla^{\mu}\, T_{\mu\nu}^{(\phi)} = 0\,.
\label{Tcon2}
\ee
Multiplying Eq.~\eqref{Tcon2} by $u_c^\nu$ and using Eq.~\eqref{Tcon}, we obtain
\be
u_c^\nu\, \nabla^\mu T^{(\phi)}_{\mu\nu} = - u_c^{\nu}\, \nabla^{\mu}\((1 + f_1) T^{(c)}_{\mu\nu}\).
\label{consTp}
\ee
The effective energy-momentum tensor for CDM, which incorporates the effect of energy exchange between the dark sectors, can be defined as
\be
\tilde{T}^{(c)}_{\mu\nu} \equiv (1+f_1)\, T^{(c)}_{\mu\nu}.
\label{tildeT}
\ee
Due to the conservation equation $\nabla^\mu (\tilde{T}^{(c)}_{\mu\nu} + T^{(\phi)}_{\mu\nu}) = 0$,
the left-hand-side of Eq.~\eqref{consTp} can be written in terms of $\tilde{T}^{(c)}_{\mu\nu}$.
Therefore we have
\be
u_c^\nu\, \nabla^\mu \tilde{T}^{(c)}_{\mu\nu} = u_c^{\nu}\, T^{(c)}_{\mu\nu}\, \nabla^{\mu} f_1,
\label{consCDMTp}
\ee
where $\nabla^\mu T^{(c)}_{\mu\nu} = 0$ has been used.

\subsection{Scaling Lagrangian}
\label{sec2.2}

The scaling solution is an interesting feature of dark energy in the context of background evolution.
The conditions for the scaling solutions can be found by analyzing evolution equations for the background universe.
The line element of the spatially flat Friedmann-Lema\^itre-Robertson-Walker (FLRW) universe is given by 
\be
ds^2 = - dt^2 + a(t)^2 \,\delta_{ij}\, dx^i\, dx^j\,,
\ee
where $\delta_{ij}$ is the Kronecker delta, and $a(t)$ is the scale factor.
The Friedmann and acceleration equations can be derived from Eq.~\eqref{Ein} as
\ba
& & 
3M_{\rm pl}^2 \,H^2 =
\rho_r + \rho_b + \tilde{\rho}_c + \rho_{\phi}\,,
\label{Eq00}\\
& & 
M_{\rm pl}^2 \left( 2\dot{H} + 3 H^2 \right)
= - P_r - P_{\phi}\,,
\label{Eq11}
\ea
where $H \equiv \dot{a} / a$ is the Hubble parameter, with a dot denoting a derivative with respect to time.
$P_r$ and $P_\phi$ are the pressures of radiation and the scalar field, respectively.
The definition Eq.~\eqref{tildeT} yields
\be
\tilde{\rho}_c \equiv \left( 1+f_1 \right) \rho_c\,.
\ee
The conservation equation \eqref{Tcon} for matter components gives
\be
\dot{\rho}_m + 3H \left( \rho_m + P_m \right) = 0\,,
\quad\mbox{for}\quad m = r, b, c\,.
\label{con}
\ee
The evolution equation for $\tilde{\rho}_c$ is obtained from Eq.~\eqref{consCDMTp} as
\be
\dot{\tilde{\rho}}_c + 3H\tilde{\rho}_c = \frac{\dot{f}_1}{1+f_1}\,\tilde{\rho}_c\,.
\label{con2}
\ee
It follows from Eq.~\eqref{Tcon2} that
\be
\dot{\rho}_{\phi} + 3 H \left( \rho_{\phi} + P_{\phi} \right)= - \frac{\dot{f}_1}{1+f_1}\,\tilde{\rho}_c\,.
\label{Eq18} 
\ee
The right-hand sides of Eqs.~\eqref{con2} and \eqref{Eq18} correspond to the energy exchange between scalar field $\phi$ and CDM.
From Eq.~\eqref{Tmm-phi}, the energy density and pressure of the scalar field take the form
\ba
\rho_{\phi} &\equiv& - \rho_c f_{1,X}\,\dot{\phi}^2 
- \rho_c f_{1,Z}\, \dot{\phi} - f_2 + f_{2,X}\, \dot{\phi}^2
+ f_{2,Z}\, \dot{\phi}\,,
\label{rhophi}\\
P_{\phi} &\equiv& f_2\,,
\label{pphi}
\ea
where the subscripts ${}_{, X}$ and ${}_{, Z}$ denote derivative with respect to $X$ and $Z$, respectively.
The equation of state of the scalar field is
\be
w_{\phi} \equiv \frac{P_{\phi}}{\rho_{\phi}}\,.
\ee
The scaling solution can be defined as a solution that satisfies
\be
\frac{\rho_\phi}{\tilde{\rho}_c} = \mbox{constant}\,.
\label{dphiH}
\ee
Since the scaling solution arises due to the coupling between the scalar field and CDM,
the scaling Lagrangian can be obtained by assuming that the main contributions to the background dynamics come from the scalar field and CDM.
When the scaling solution is achieved,
Eq.~\eqref{Eq00} gives $\rho_\phi \propto (1 + f_1) \rho_c \propto H^2$.
Assuming that $w_\phi$ is constant during the scaling regime, $P_\phi \propto \rho_\phi$. 
consequently Eq.~\eqref{pphi} yields $f_2 \propto H^2$,
and each term on the left-hand side of Eq.~\eqref{Eq18} is proportional to $H^3$.
Hence, $\dot{f}_1 \propto H^3 (1 + f_1) / \tilde\rho_c \propto H (1 + f_1)$. 
From these relations, the expressions for $f_1$ and $f_2$, derived in \cite{scaling}, can be expressed as 
\begin{eqnarray}
f_1(\phi, X, Z) &=& e^{Q\phi/M_{\rm pl}}\, g_1(Y_1, Y_2) - 1\, ,
\label{eq:simpY1}\\
f_2(\phi, X, Z) &=& X\, g_2(Y_1, Y_2)\,,
\label{eq:simpY2}
\end{eqnarray}
where $Q$ is a constant, while $g_1(Y_1,Y_2)$ and $g_2(Y_1,Y_2)$ are arbitrary functions of $Y_1$ and $Y_2$, defined as
\begin{eqnarray}
Y_1 &=& X\, e^{\lambda \phi/M_{\rm pl}} \,,
\label{eq:Y1}\\
Y_2 &=& Z\, e^{\lambda \phi/(2 M_{\rm pl})}\,,
\label{eq:Y2}
\end{eqnarray}
where $\lambda$ is a constant parameter.
Inserting Eqs.~\eqref{eq:simpY1} and \eqref{eq:simpY2} into Eq.~\eqref{Lin},
we obtain
\be
L = -\left[ e^{Q\phi/M_{\rm pl}}\, g_1 \left(Y_1,Y_2 \right)-1 \right]\, \rho_c + X\, g_2 (Y_1, Y_2)\,.
\label{slag}
\ee
To ensure that $g_1$ and $g_2$ remain finite when $M_{\rm pl} e^{-\lambda \phi/(2M_{\rm pl})} / (\sqrt{3}H) \to 0$ during $\phi$MDE,
the functions $g_1$ and $g_2$ can be constructed in polynomial form as \cite {scaling}
\begin{eqnarray}
\label{g1}
g_1(Y_1,Y_2) &=& b_0 + \sum_{i>0}{(b_i\, Y_1^{-i} + \tilde{b}_i\, Y_2^{-i})} + 2^{1-m/2}\,\mu\, \frac{Y_2^m}{Y_1^{m/2}} + \sum_{i>0,j<2i}{\mu_i\, \frac{Y_2^j}{Y_1^i}}\,,
\\ 
\label{g2}
g_2(Y_1,Y_2) &=& c_0 + \sum_{i>0}{(c_i\, Y_1^{-i} + \tilde{c}_i\, Y_2^{-i})} + 2^{1-m/2}\, \beta\, \frac{Y_2^m}{Y_1^{m/2}} + \sum_{i>0,j<2i}{\beta_i\, \frac{Y_2^j}{Y_1^i}}\,,
\end{eqnarray}
where $b_0, b_i, \tilde{b}_i, \mu, \mu_i, c_0, c_i, \tilde{c}_i, \beta, \beta_i$ and $m$ are constants.
In our analysis,
the form of $g_1$ is chosen such that the dark matter energy density is coupled to the scalar field through the exponential factor $e^{Q\phi/M_{\rm pl}}$ in Eq.~\eqref{slag}. 
Since $f_2$ is equal to $P_\phi$, the form $g_2$ is chosen such that $f_2$ reduces to the canonical form $f_2 = X - V$ when the coupling vanishes.
Based on Eqs.~\eqref{g1} and \eqref{g2}, these requirements can be satisfied if
\begin{eqnarray}
g_1(Y_1,Y_2) &=& 1\,,
\label{eq:simpg1}\\
g_2(Y_1,Y_2) &=& 1-\frac{V_0}{Y_1}+2^{1-m/2}\, \beta\, \frac{Y_2^m}{Y_1^{m/2}}\,.
\label{eq:simpg2}
\end{eqnarray}
Applying $g_1$ and $g_2$ to Eqs.~\eqref{eq:simpY1} and \eqref{eq:simpY2}, we get
\be
f_1 = e^{Q\phi/M_{\rm pl}} - 1\,,
\quad
f_2 = X - V_0\, e^{-\lambda \phi/M_{\rm pl}} + \beta\,(2 X)^{1 - m/2}\, Z^m \,.
\label{eq:f1f2}
\ee
Therefore, the scaling Lagrangian Eq.~\eqref{slag} can be demonstrated as 
\be
L = - \left( e^{Q\phi/M_{\rm pl}} - 1 \right)\, \rho_c (n_c) + \left( X - V(\phi) \right) + \beta\,(2 X)^{1 - m/2}\, Z^m \,,
\label{eq:scaling lagrangian}
\ee
where $V(\phi) = V_0\, e^{-\lambda\phi / M_{\rm pl}}$, $\lambda$ is a constant, $V_0$ is a positive constant, $Q$ and $\beta$ are constant parameters that characterize the energy and momentum exchange between CDM and scalar field, respectively.
The dynamics of the background universe and growth of the linear matter perturbations on small scales for dark energy with the Lagrangian in Eq.~\eqref{eq:scaling lagrangian} have been investigated in \cite{scaling}.
This form of Lagrangian can provide the scaling $\phi$-Matter Dominated Epoch ($\phi$MDE), followed by another scaling epoch describing the cosmic acceleration that is consistent with observational constraints.
The growth of CDM perturbations on small scales can be suppressed compared with the $\Lambda$CDM model, potentially alleviating $\sigma_8$ tension.
Before concluding this section, we discuss some physical consequences of the energy and momentum couplings. 
It follows from Eqs.~\eqref{con2} and \eqref{Eq18} that the momentum coupling does not affect the background evolution of the energy density. 
For the choice of $f_1$ given in Eq.~\eqref{eq:f1f2}, 
Eq.~\eqref{con2} becomes
 \be
\dot{\tilde{\rho}}_c = - 3H\tilde{\rho}_c + \frac{Q}{M_{\rm pl}}\dot\phi \tilde{\rho}_c\,.
\label{dotrhoc}
\ee
For runaway quintessence, $\dot\phi>0$, 
so the energy coupling enhances the decay rate of $\tilde{\rho}_c$ when $Q<0$, 
indicating that the background energy of CDM is transferred to dark energy. 
Equation~\eqref{dotrhoc} can be integrated to give
\be
\tilde{\rho}_c = \tilde{\rho}_c^0 a^{-3} e^{Q (\phi - \phi^0) / M_{\rm pl}}\,,
\ee
where the superscript ${}^0$ denotes the present-day value. 
Hence, if $|Q|$ is large, the evolution of $\tilde{\rho}_c$ can deviate significantly from its evolution in the uncoupled case. 
As a result, the growth of matter perturbations during the matter-dominated epoch is substantially modified, allowing observations of the large-scale structure of the Universe to place constraints on the magnitude of $Q$. 
The influence of the momentum coupling can be investigated through the evolution equations of matter perturbations on small scales, which are considered in the next section.

\section{Spherical Collapse}
\label{sec3}

To investigate spherical collapse, we compute the evolution equation for matter perturbations up to second order on small scales, 
where the time derivatives of perturbed variables can be neglected compared with their spatial derivatives under the quasi-static approximation \cite{operatorlambda}.
In coupled dark energy models, baryons and CDM collapse at different rates due to the coupling between dark energy and CDM, causing their perturbations to reach the turnaround radius at different times. 
Consequently, baryons flow out of the CDM top-hat region. 
In this situation, the virial mass of the collapsing object and the extrapolated linear density contrast at collapse must be defined with care \cite{Mainini:2005fe, Mainini:2006zj}. 
In this work, we simplify the analysis by treating baryons and CDM as a single fluid coupled to dark energy in the spherical collapse model. 
A brief comparison between the results obtained under this simplification and those obtained in \cite{Mainini:2006zj} is presented in Appendix~\ref{compare}. 
This comparison shows that the cluster number density predicted in our simplified treatment differs from that of \cite{Mainini:2006zj} by less than 10\%\,.

The perturbed spacetime is described by the following line element: 
\be
ds^2 = - (1 - 2 \Phi)\, dt^2 + a^2\, (1 + 2\Phi)\, \delta_{ij}\, dx^i dx^j\,,
\label{permet}
\ee
where $\Phi$ is the metric perturbation in the Newtonian gauge, and anisotropic stress is assumed to vanish.
The evolution equation for the matter density contrast $\delta_m$ is obtained by expanding the $\nu=0$ component of Eq.~\eqref{Tcon} up to second order:
\be
\dot{\delta}_m + (1 + \delta_m) u^{i}_{,i} + \delta_{m,i} u^i = 0\,, 
\label{dotdel1}
\ee
where the subscript ${}_{,i}$ denotes a derivative with respect to the spatial comoving coordinate $x^i$, $u^i$ is the velocity of cold dark matter, 
and $\delta_m \equiv \delta\rho_m / \bar\rho_m$, where $\delta\rho_m$ and $\bar\rho_m$ denote the matter density perturbation and the background energy density, respectively.
Expanding the $\nu = i$ component of Eq.~\eqref{Tcon2} up to second order in perturbations, we obtain
\ba
\label{dv1}
&& \(1 + \delta_m + \frac{2 \beta m x^2}{\Omega_m} \) \dot{u}_{i} 
+ \frac{H[\Omega_m (2 q_s - 2 \sqrt{6}\, \beta m Q x + \sqrt{6}\, Q q_s x) + 2\beta mx(\sqrt{6}\,\lambda y^2 - q_s x)]}{q_s \Omega_m}u_{i} \nonumber\\
&& + \bigl(\dot{\delta}_m + \delta_m H (2 + \sqrt{6}\, Q x) + u_{j}^{,j}\bigr) u_{i} + u_{i ,j} u^{j} - \frac{(1 + \delta_m)}{a^2}\Phi_{,i}
\nonumber\\
&&
+ \frac{\sqrt{6}\, \beta m q_s x}{3 q_s a^2 H \Omega_m}\delta\dot{\phi}_{,i} + \frac{Q \Omega_m\bigl(1 + \beta (2 - m)\bigr) + \lambda \bigl(q_s - 1 - \beta (2 - m)\bigr) y^2}{q_s a^2 \Omega_m} \delta{\phi}_{,i}
= 0\,.
\ea
Here, $\delta\phi$ is the perturbation of the scalar field $\phi$, and the dimensionless variables are defined similarly to those in \cite{scaling}; their expressions are given in Appendix~\ref{appen1}.
Inserting $\dot{\delta}_m$ from Eq.~\eqref{dotdel1} into Eq.~\eqref{dv1}, we obtain
\ba
\label{dv2}
&&\dot{u}_{i} + \frac{H \left[q_s\Omega_m (1 + \delta_m) \left(2 + \sqrt{6}\, Q x\right) + 2\beta m x \left(\sqrt{6}\, \lambda y^2 - \sqrt{6}\,Q \Omega_m - q_s x\right) \right]}{q_s(\Omega_m (1 + \delta_m) + 2 \beta m x^2)}\,u_{i} \nonumber\\
&&- \frac{\Omega_m (\delta_m u_{i} u_{j}^{,j} + u_{i} u_{j} \delta_m^{,j} - u_{i,j} u^{j})}{\Omega_m (1 + \delta_m) + 2 \beta m x^2} - \frac{\Omega_m (1 + \delta_m)}{a^2 [\Omega_m (1 + \delta_m) + 2 \beta m x^2]}\, \Phi_{,i} \nonumber\\
&&+ \frac{\sqrt{6}\, \beta m x}{3 a^2 H [\Omega_m (1 + \delta_m) + 2 \beta m x^2]}\,\delta\dot{\phi}_{,i} + \frac{Q \Omega_m [1 + \beta (2 - m)] + \lambda [q_s - 1 + \beta (m - 2)] y^2}{q_s a^2 [\Omega_m (1 + \delta_m) + 2 \beta m x^2]}\, \delta\phi_{,i} = 0\,.
\ea
Differentiating Eq.~\eqref{dotdel1} with respect to time, we obtain
\begin{align}
\label{ddotdelc1}
&\ddot{\delta}_m + \frac{H \bigg(-2 \beta m q_s x^2 + q_s \Omega_m (2 + \sqrt{6}\, Q x) + 2 \sqrt{6}\, \beta m x (- Q \Omega_m + \lambda y^2)\bigg) \ \dot{\delta}_m}{q_s (\Omega_m + 2 \beta m x^2)} 
- \frac{\dot{\delta}_m^2}{1 + \delta_m} \nonumber\\ 
&+\frac{\Bigl(\bigl(-1 + \beta (-2 + m)\bigr) Q \Omega_m - \lambda \bigl(-1 + \beta (-2 + m) + q_s\bigr) y^2\Bigr)(1 + \delta_m) \delta\phi^{,i}_{,i} }{q_s a^2 (\Omega_m + 2 \beta m x^2)} \nonumber\\
&- \frac{\sqrt{2}\, \beta m (1 + \delta_m) x \delta\dot{\phi}^{,i}_{,i}}{\sqrt{3}\, a^2 H (\Omega_m + 2 \beta m x^2)} + \frac{(1 + \delta_m) \Omega_m \Phi^{,i}_{,i}}{a^2 (\Omega_m + 2 \beta m x^2)}
- \frac{\Omega_m (u^i u^j_{,ji} + u_{j, i} u^{j, i} ) (1+\delta_m)}{\Omega_m + 2 \beta m x^2} = 0\,,
\end{align}
where $\dot{u}^i_{,i}$ is eliminated using Eq.~\eqref{dv2}, 
while $u^i_{,i}$ is replaced using $u^i_{,i} = - \dot{\delta}_m/(1 + \delta_m)$ from Eq.~\eqref{dotdel1}. 
We also adopt the top-hat assumption $\delta_{m,i} = 0$.
Using the identity
\be
u^i\, \partial_i \partial_j u^j = \frac 13 (\partial_i u^i)^2 - \partial_i u_{j}\, \partial^i u^j\,,
\ee
Eq.~\eqref{ddotdelc1} becomes
\begin{align}
\label{ddotdelc2}
&\ddot{\delta}_m + \frac{\bigl(-2 \beta m q_s x^2 + q_s \Omega_m (2 + \sqrt{6}\, Q x) + 2 \sqrt{6}\, \beta m x (- Q \Omega_m + \lambda y^2)\bigr) H \dot{\delta}_m}{q_s (\Omega_m + 2 \beta m x^2)} 
- \frac{2 (2 \Omega_m + 3 \beta m x^2) \ \dot{\delta}_m^2}{3 (1 + \delta_m) (\Omega_m + 2 \beta m x^2)} \nonumber\\
& + \frac{ \Bigl(\bigl(-1 + \beta (-2 + m)\bigr) Q \Omega_m -  \lambda \bigl(-1 + \beta (-2 + m) + q_s\bigr) y^2\Bigr) (1 + \delta_m) \delta\phi^{,i}_{,i} }{q_s a^2 (\Omega_m + 2 \beta m x^2)}\nonumber\\
&- \frac{\sqrt{2}\, \beta m x (1 + \delta_m) \delta\dot{\phi}^{,i}_{,i} }{\sqrt{3}\, a^2 H (\Omega_m + 2 \beta m x^2)}
+ \frac{\Omega_m (1 + \delta_m) \Phi^{,i}_{,i}}{a^2 (\Omega_m + 2 \beta m x^2)} = 0\,.
\end{align}
On small scales, the $(00)$ component of the Einstein equations leads to
\be
\Phi_{, i}^{, i} = \frac{3}{2} a^2 H^2 \Omega_m \delta_m \,,
\label{Phiii}
\ee
while the $\mu = 0$ component of Eq.~\eqref{Tcon2} gives
\be
\delta\phi_{, i}^{, i} =
- \frac{a^2 H^2}{2 c_s^2 q_s x} \(\sqrt{6}\, Q \Omega_m \delta_m - 2 \beta m x \dot\delta_m\)\,.
\label{Piii}
\ee
Inserting Eqs.~\eqref{epsilonbg}, \eqref{xibg}, \eqref{Piii}, \eqref{Phiii}, and the time derivative of Eq.~\eqref{Piii} into Eq.~\eqref{ddotdelc2}, we obtain
\be
\label{Non-Linear-dddeltaC}
\delta_m'' + \nu \delta_m' + \mu_1 \frac{ \delta_m'^2}{1+\delta_m} - \frac{3}{2} G_{cc} \Omega_m \delta_m (1 + \delta_m) = 0\,,
\ee
where we have changed the time derivative to a derivative with respect to $N$.
The expressions for $\nu$ and $G_{cc}$ agree with those in the linear evolution equation for $\delta_m$ in \cite{scaling} and are given in Appendix~\ref{apx2}. 
The coefficient $\mu_1$ is
\be
\mu_1 = \frac{\bigl(4 -  \beta (-8 + m)\bigr) \Omega_m + 6 \beta (1 + 2 \beta) m x^2}{3 \bigl(-1 + \beta (-2 + m)\bigr) \Omega_m - 6 \beta (1 + 2 \beta) m x^2}\,.
\ee
In the numerical integration, the initial conditions for the density contrast can be specified by considering its behavior during the matter-dominated epoch. 
During this epoch, the density contrast is initially small; therefore, the nonlinear terms can be neglected. 
Equation~\eqref{Non-Linear-dddeltaC} then reduces to
\be
\delta_{m}'' + \frac 12 \delta_{m}' - \frac{3 + 6 \beta + 6 Q^2}{2 + 4 \beta} = 0 \,.
\label{Linear-dddeltaC}
\ee
The solution of Eq.~\eqref{Linear-dddeltaC} is
\be
\delta_{m} = C_1 e^{- r N} + C_2 e^{ r N}\,,
\label{delta_early}
\ee
where $C_1$ and $C_2$ are integration constants.
We choose the growing solution by setting $C_1 = 0$ and $C_2 = \delta_{i}$, where
\be
r = \frac14\, \Big[-1 + \sqrt{25+ 48 \frac{Q^2}{1 + 2\beta}}\Big]\,.
\label{exponent}
\ee
Hence, the initial conditions for Eq.~\eqref{Non-Linear-dddeltaC} in the matter-dominated epoch can be written as
\be
\delta_{m}|_{t = t_i} = \delta_i\,,
\quad
\delta_{m}' |_{t = t_i} = r \delta_i\,,
\label{initialconds}
\ee
where $t_i$ is the initial time.
We now consider the linear growth factor, defined as
\be
D(z) \equiv \frac{\delta_m(z)}{\delta_m(z = 0)}\,,
\ee
where $\delta_m(z)$ is the linear density contrast obtained from the linearized version of Eq.~\eqref{Non-Linear-dddeltaC} using the initial conditions in Eq.~\eqref{initialconds}.
In \tref{tab:1}, we present the parameter values for each model. 
The parameters $\lambda$, $Q$, $\beta$, and $m$ are constrained by datasets from Planck 2018, the Pantheon supernova sample, and the first-year DES results, as analyzed in \cite{bestfit}.
Since our focus is on the impact of momentum coupling, the parameters $\lambda$, $Q$, and $m$ are set identical across all models and are fixed to the mean values reported in \cite{bestfit}, 
while the strength of the momentum transfer, $\beta$ is varied.
The value of $\beta$ in Model A is set to its mean value, while Models B and C take values corresponding to the $1\sigma$ and $2\sigma$ confidence levels, respectively.
However, to compare with the effects of energy coupling, we also consider Model D, in which the coupling magnitude $Q$ is increased to the $1\sigma$ confidence level.
The influences of energy and momentum couplings on the growth of matter perturbations can be understood from Eq.~\eqref{exponent}.
This equation indicates that, during the matter-dominated epoch, the growth rate of $\delta_m$ is enhanced by the parameter $Q$, while, for a given value of $Q$, it is suppressed by the parameter $\beta$.
 Although the sign of $Q$ determines the direction of energy transfer between dark energy and CDM, 
 the growth rate of matter perturbations does not depend on it. 
 It follows from Eq.~\eqref{exponent} that, if $Q$ vanishes, the momentum coupling does not affect the growth rate of matter perturbations during the matter-dominated epoch. 
 However, when the dark energy density becomes significant, $\nu$ and $G_{cc}$, given by Eqs.~\eqref{coefnu} and \eqref{def:gcc}, do not vanish even if $Q=0$. 
 This indicates that the momentum coupling can affect the growth rate of matter perturbations once dark energy makes a significant contribution to the background dynamics, even in the absence of energy coupling. 
 Hence, the growth of matter perturbations is influenced by the energy and momentum couplings at different cosmic epochs, and can therefore be probed through observations of the large-scale structure of the Universe.

\begin{table}[width=0.9 \linewidth,cols=4,pos=h]
\caption{Values of parameters $\lambda, Q, \beta$ and $m$ for each model.}
\label{tab:1}
\begin{tabular*}{\tblwidth}{@{} L|LLLL@{} }
\toprule
Parameters & Model A & Model B & Model C & Model D\\
\midrule
$\lambda$ & 0.4042 & 0.4042 & 0.4042 & 0.4042 \\
Q & -0.0312 & -0.0312 & -0.0312 & -0.0624 \\
$\beta$ &  0.332 & 0.095 & 0.011 & 0.332 \\
m & 2 & 2 & 2 & 2 \\
\bottomrule
\end{tabular*}
\end{table}

\begin{figure}
\centering
\includegraphics[height=0.4\textwidth, width=0.6\textwidth, angle=0]{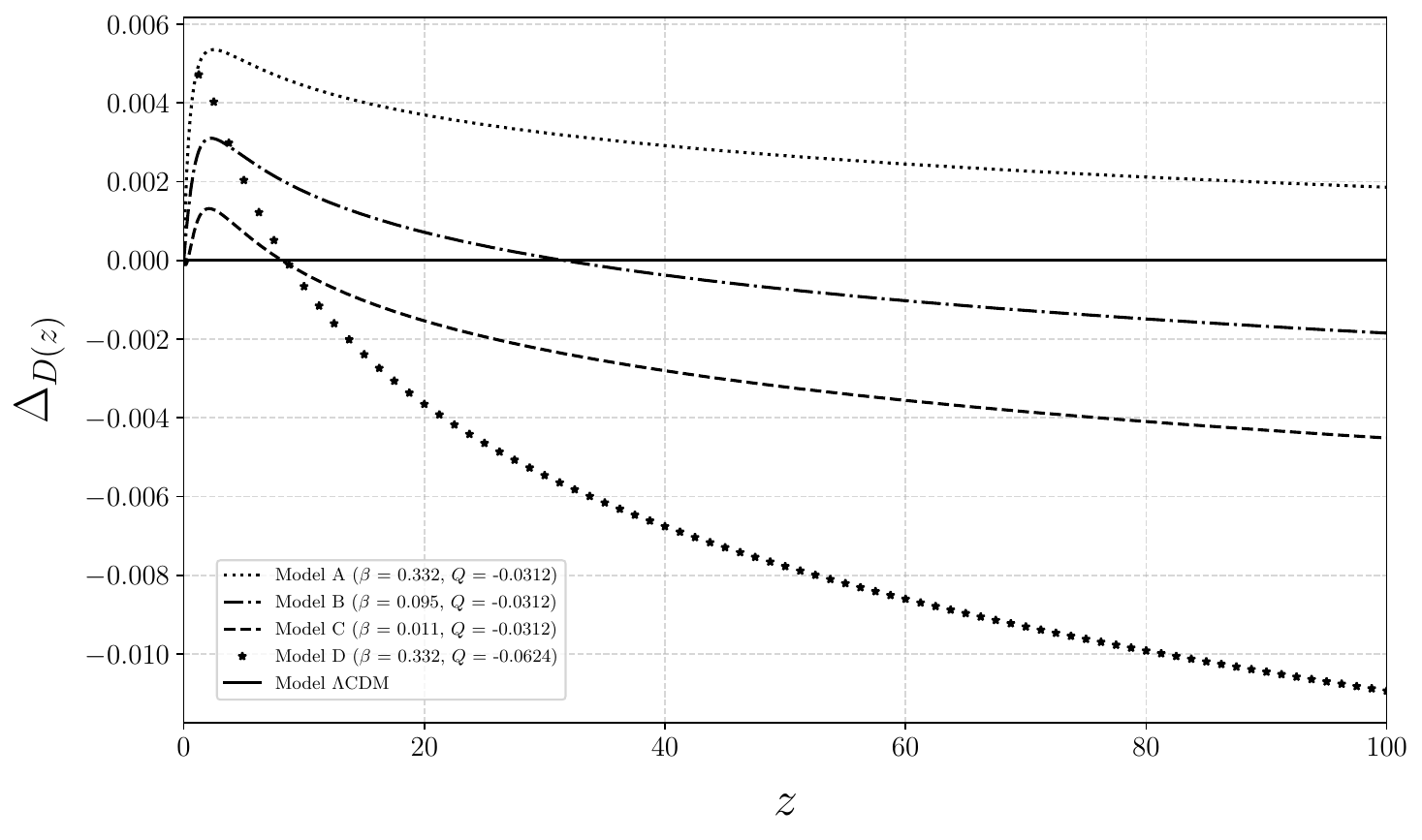}
\caption{Plot of difference ratio $\Delta_{D(z)}$. The lines A, B, C and D represent the different ratios of the models A, B, C and D with the $\Lambda$CDM model.}
\label{diff_ratio}
\end{figure}

In Fig.~\ref{diff_ratio}, we plot the difference ratio $\Delta_{D(z)} \equiv (D(z)/D(z)_{\Lambda})- 1$ for Models A, B, C and D, where $D(z)_{\Lambda}$ is the linear growth factor for the $\Lambda$CDM model.
The difference ratio $\Delta_{D(z)}$ gradually increases as $z$ decreases for $z < 100$.
This ratio reaches its maximum at $z\approx 0.3$ for all coupled models.
For $z < 0.3$, $\Delta_{D(z)}$ decreases toward unity at the present. 
While the difference ratio $\Delta_{D(z)}$ for Model A is positive throughout the entire history of the Universe, those for Models B and C are positive at $z<30$ and $z<10$, respectively.
The positivity of the difference ratio implies that the growth of matter perturbations is slower than in the $\Lambda$CDM model, as the growth factor is normalized to unity at present.
It follows from~\fref{diff_ratio} that the difference ratio for Model D is smaller than that for Model A over a wide range of redshift, 
suggesting that the energy coupling enhances the growth rate of matter perturbations.

\begin{figure}
\centering
\includegraphics[height=0.4\textwidth, width=0.6\textwidth]{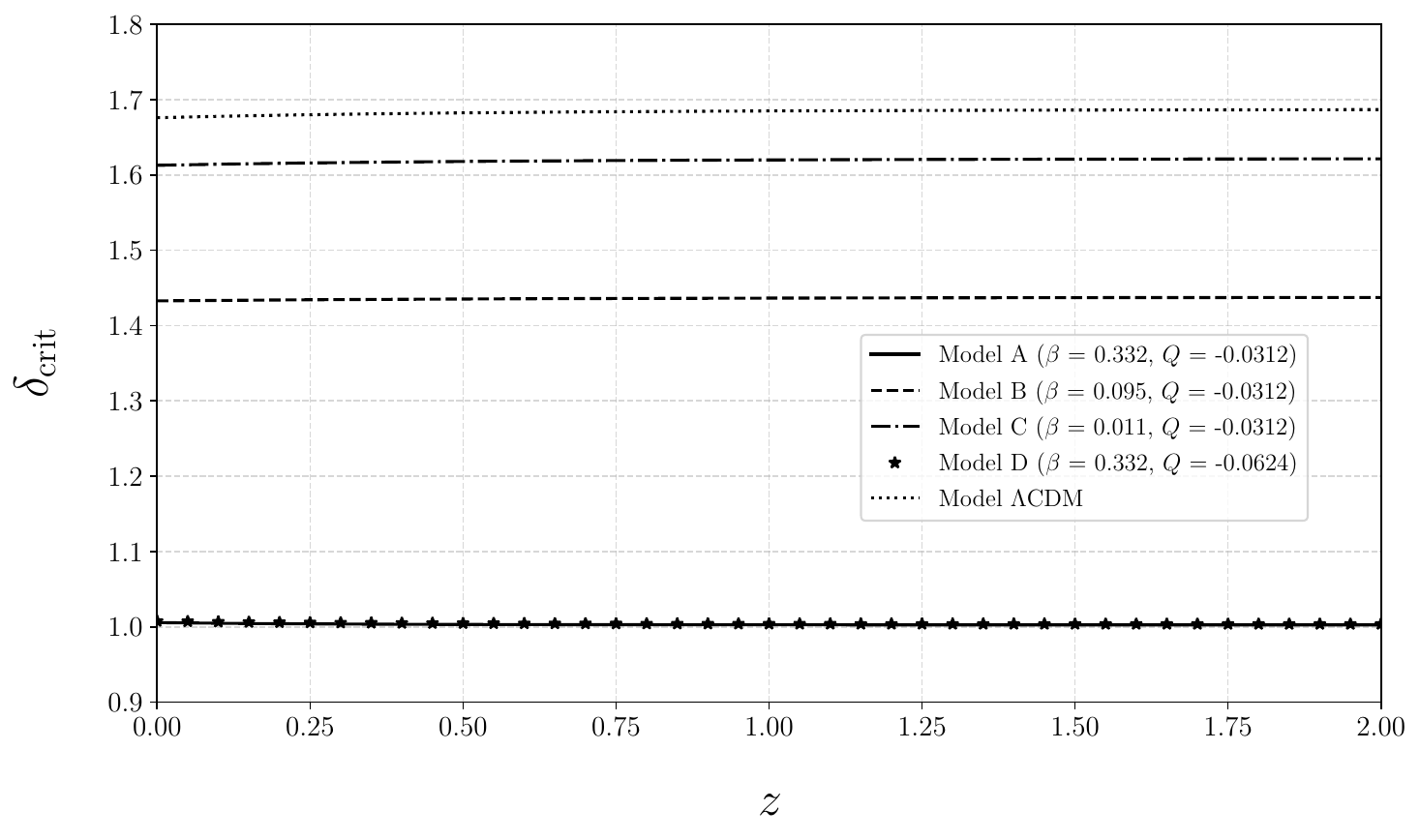}
\caption{Plot of linear density contrast at collapse $\delta_\crit$ as a function of collapsed redshift $z$. The lines A, B, C, D and $\Lambda$CDM correspond to models A, B, C, D and $\Lambda$CDM. }
\label{delC_compare}
\end{figure}

One of the important results of the spherical collapse model is the extrapolated linear density contrast $\delta_{\rm crit}$, 
which is a parameter used in the calculation of the mass function.
This quantity is computed by numerically solving Eq.~\eqref{Non-Linear-dddeltaC} and determining the initial conditions for $\delta_m$ that lead to the collapse of an overdense region, 
i.e.,\ $\delta_m \to \infty$, at a given redshift.
Using these initial conditions for $\delta_m$ in Eq.~\eqref{Linear-dddeltaC}, the extrapolated linear density contrast is obtained by evolving $\delta_m$ linearly to the collapse redshift.
In our calculation, we start at a redshift where $\delta_m$ obeys the linear evolution equation, allowing us to set the initial conditions for $\delta_m$ and $\delta_m'$ according to Eq.~\eqref{initialconds}.
Plots of $\delta_{\rm crit}(z)$ for Models A, B, C, D and the $\Lambda$CDM are shown in~\fref{delC_compare}. 
As seen from the figure, $\delta_{\rm crit}(z)$ decreases as the strength of the momentum coupling $\beta$ increases, implying that momentum coupling facilitates the collapse.
Increasing the strength of the energy coupling in Model D does not significantly affect $\delta_{\rm crit}(z)$ compared with Model A.
\section{Cluster Number Counts}
\label{sec4}

The comoving number density of collapsed objects with masses between $M$ and $M+dM$
can be estimated using the differential halo mass function, $n(M, z)$, and the extrapolated linear density contrast from the spherical collapse model, as proposed by \cite{Press-Schechter}:
\be
n(M, z)dM =
- \frac{\hat{\rho}_m}{M^2}\frac{d \ln \sigma(M, z)}{d \ln M} f(\sigma)\, dM,
\label{diffrel}
\ee
where $\hat{\rho}_m \equiv a^3 \tilde{\rho}_m$ is the comoving energy density of matter.
$\sigma(R(M), z)$ is the root-mean-square amplitude of the linear matter density variance in a region with comoving radius $R$ enclosing a mass $M$ at redshift $z$, 
which can be computed from the power spectrum as \cite{numbercount2}
\be
\sigma^2(R(M), z) = \frac{1}{2\pi^2}\int_0^\infty k^2 P(k, z) W^2(k R(M))\, d k\,.
\label{varinf}
\ee 
The linear matter power spectrum for the coupled dark energy model at a given redshift $P(k, z)$ is computed using a modified version of CLASS, 
with the cosmological parameters set according to the best-fit values in~\cite{bestfit},
while that for the $\Lambda$CDM model is computed using the cosmological parameters from the observational constraints in~\cite{Riess:2018}.
The Fourier-transformed window function of a spherical top-hat function is defined as
\be
\label{eq:window_fn}
W(k R) = \frac{3}{(k R)^3} \Big(\sin(kR) - k R \cos(kR) \Big),
\ee
which serves as a smoothing filter for the density field. 
The comoving radius which represents the radius of the background region enclosing mass $M$ is given by
\be
R \equiv \bigg(\frac{3 M}{4 \pi \hat{\rho}_m} \bigg)^{1/3}\,.
\ee
The mass function $f(\sigma)$ is defined as the fraction of mass in collapsed regions per unit interval in $\ln{\sigma}^{-1}$.
Since no mass functions have been specifically developed for coupled dark energy models, 
we investigate the effects of the coupling between dark energy and dark matter on cluster number counts using the available analytic mass functions. 
In our analysis, we consider both the Press-Schechter (PS) and Sheth-Tormen (ST) mass functions. 
The ST mass function is given by \cite{Sheth-Tormen,massfunction}
\be
f_{\rm S-T}(\sigma) = A\sqrt{\frac{2a}{\pi}}\bigg[1+\bigg(\frac{\sigma^2}{a\delta_\crit^2}\bigg)^p\bigg]\frac{\delta_\crit}{\sigma}\, \text{exp}\Big[-\frac{a\delta_{\rm \crit}^2}{2\sigma^2}\Big]\,,
\label{st-mass}
\ee 
where $\delta_\crit$ is the linear extrapolated density contrast at collapsed redshift.
The parameter $A$ is a normalization constant equal to $ 0.3222$ for our mass limit. 
The parameters $a = 0.707$ and $ p = 0.3$ are the values according to the best fit with the $\Lambda$CDM model~\cite{Sheth-Tormen}.
By setting $a = 1$, and $p = 0$, the ST mass function reduces to the PS mass function.
\vspace{\baselineskip}
\begin{table}[width=0.9 \linewidth,cols=4,pos=h]
\caption{The value of $\sigma_8$ for model parameters given by Table~\ref{tab:1}.}
\label{tab:2}
\begin{tabular*}{\tblwidth}{@{} L|LLLLL@{} }
\toprule
Model & A & B & C & D & $\Lambda$CDM \\
\midrule
$\sigma_8$ & 0.802 & 0.808 & 0.811 & 0.822 & 0.811 \\
\bottomrule
\end{tabular*}
\end{table}
\vspace{\baselineskip}
\newline
The number of galaxy clusters in a redshift interval $\Delta z$ with mass larger than $M_{\rm min}$ can be computed by integrating Eq.~\eqref{diffrel} over the mass as
\be
\ N = f_{\rm sky} \frac{dV_e}{dz}\, \Delta z
\int_{M_{\rm min}}^\infty n(M)\, dM\,,
\label{dNdz}
\ee
where $f_{\rm sky}$ is the observed sky fraction,
$dV_e/dz \equiv 4\pi r(z)^2/H(z)$ is the comoving volume element per unit redshift, and
$r(z)$ is the comoving distance at redshift $z$. 
The volume element is evaluated at the midpoint of each redshift bins.

\begin{figure}
\centering
\includegraphics[height=0.4\textwidth, width=0.6\textwidth, angle=0]{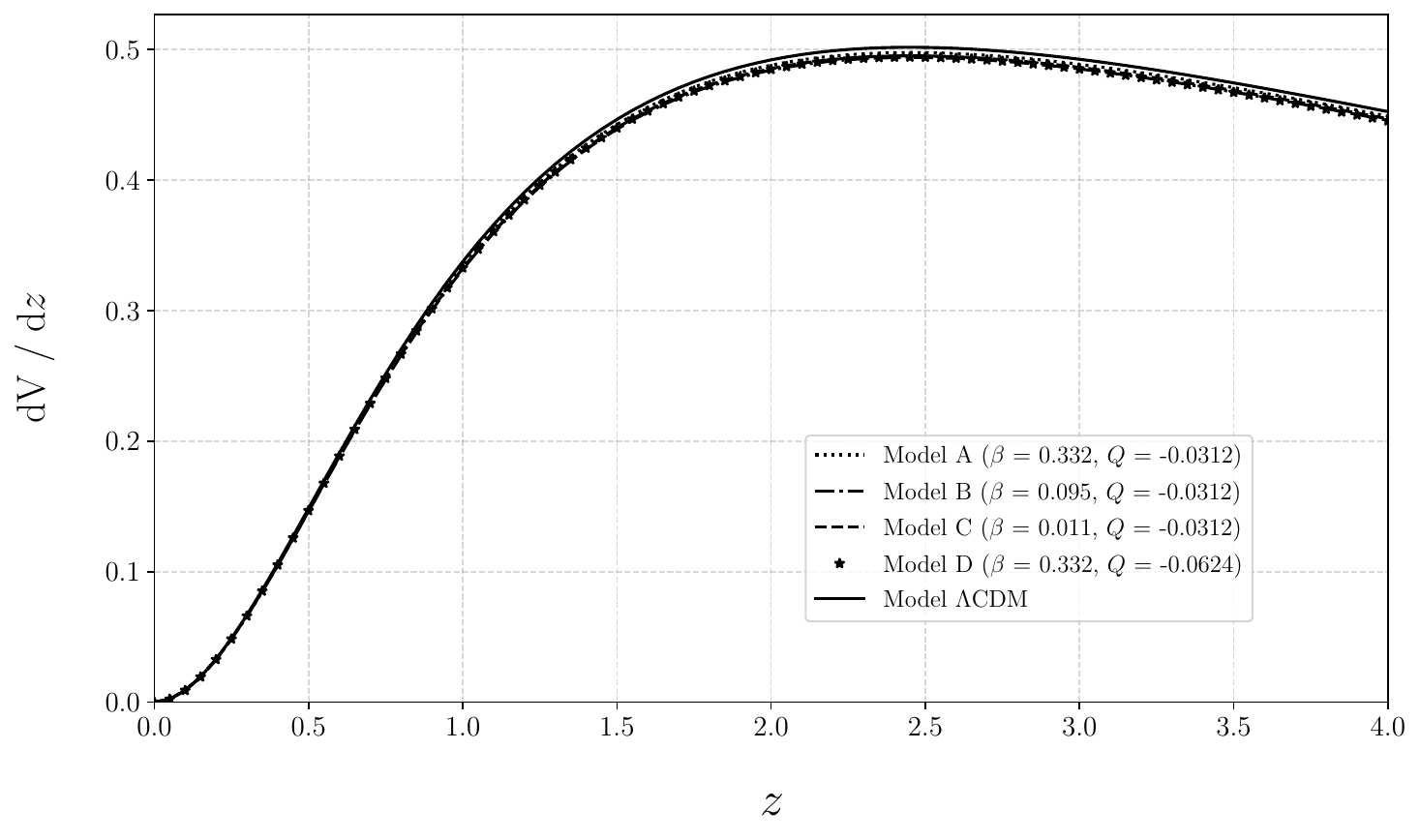}
\caption{Plots of comoving volume element $V_e$ as a function of redshift $z$. The lines A, B, C, D, and $\Lambda$CDM correspond to models A, B, C, D, and $\Lambda$CDM.}
\label{dVdz}
\end{figure}

To understand the influence of momentum coupling on the cluster number counts, 
we investigate how momentum coupling affects the comoving volume element per unit redshift $dV_e/dz$.
As shown in~\fref{dVdz}, we see that the evolution of $dV_e/dz$ for all coupled models differs only slightly from that of the $\Lambda$CDM model within the parameter ranges considered in this work.
Since the mass function depends on dynamics of matter perturbations through the ratio $\delta_{\rm crit}/(\sigma(R(M), z))$,
we investigate how the energy and momentum coupling affect this ratio by considering the case $R = 8 h^{-1}{\rm Mpc}$ for illustration.
For this case, we denote $\sigma_8(z) = \sigma(8, z)$,
so that $\sigma_8 = \sigma_8(z = 0)$.

Its values for given models are given in \tref{tab:2}.
This indicates that $\sigma_8$ decreases with increasing momentum coupling strength $\beta$, 
whereas it increases with the magnitude of the coupling constant $Q$.
\begin{figure}
\centering
\includegraphics[height=0.4\textwidth, width=0.6\textwidth, angle=0]{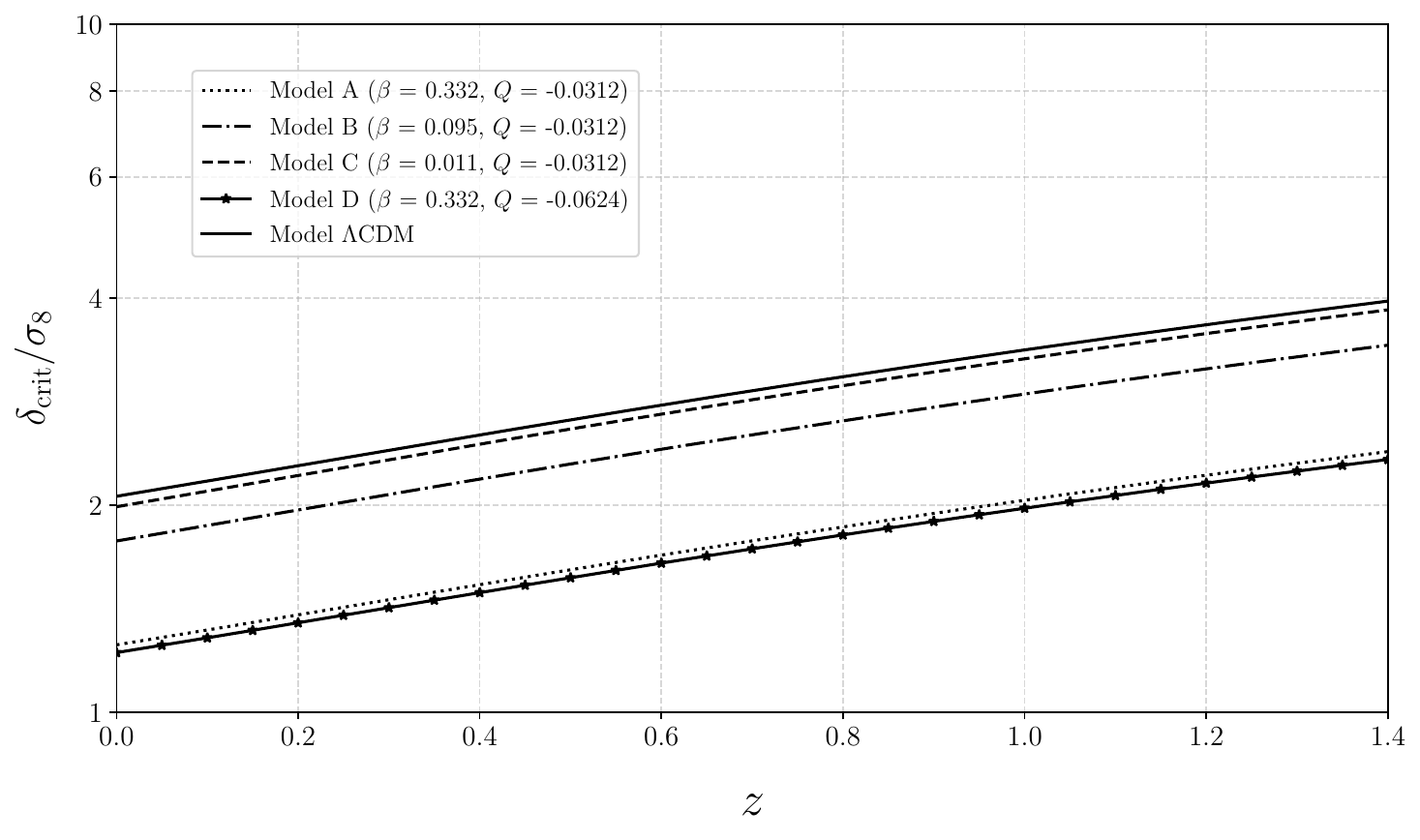}
\caption{Plots of $\delta_\crit / \sigma_8(z)$ as a function of redshift $z$. The lines A, B, C, D, and $\Lambda$CDM correspond to models A, B, C, D, and $\Lambda$CDM, in order.}
\label{delcs8d}
\end{figure}
It follows from~\fref{delcs8d} that the ratio $\delta_\crit / \sigma_8(z)$ decreases from high redshift to the present.
At a given redshift, this ratio decreases as $\beta$ increases and as the magnitude of $Q$ increases.
This can be seen from~\fref{diff_ratio} that a nonzero $\beta$ weakly enhances the growth factor $D(z)$,
while $\delta_\crit$ is significantly suppressed as shown in~\fref{delC_compare}.
Consequently, the ratio $\delta_\crit / \sigma_8(z)$ reduces when momentum coupling $\beta$ increases.
The enhancement of the magnitude of $Q$ in Model D does not significantly influence $\delta_\crit$ compared with Model A, 
but $\sigma_8$ increases significantly relative to the other models due to the enhanced growth rate of matter perturbations on small scales.
This indicates that an increase in $\beta$ or in the magnitude of $Q$ enhances the mass function in Eq.~\eqref{st-mass} by reducing the factor $\delta_{\rm \crit}^2 / \sigma^2$ in the exponential term.

\begin{figure}
\centering
\includegraphics[width=0.5\textwidth, width=0.8\textwidth, angle=0]{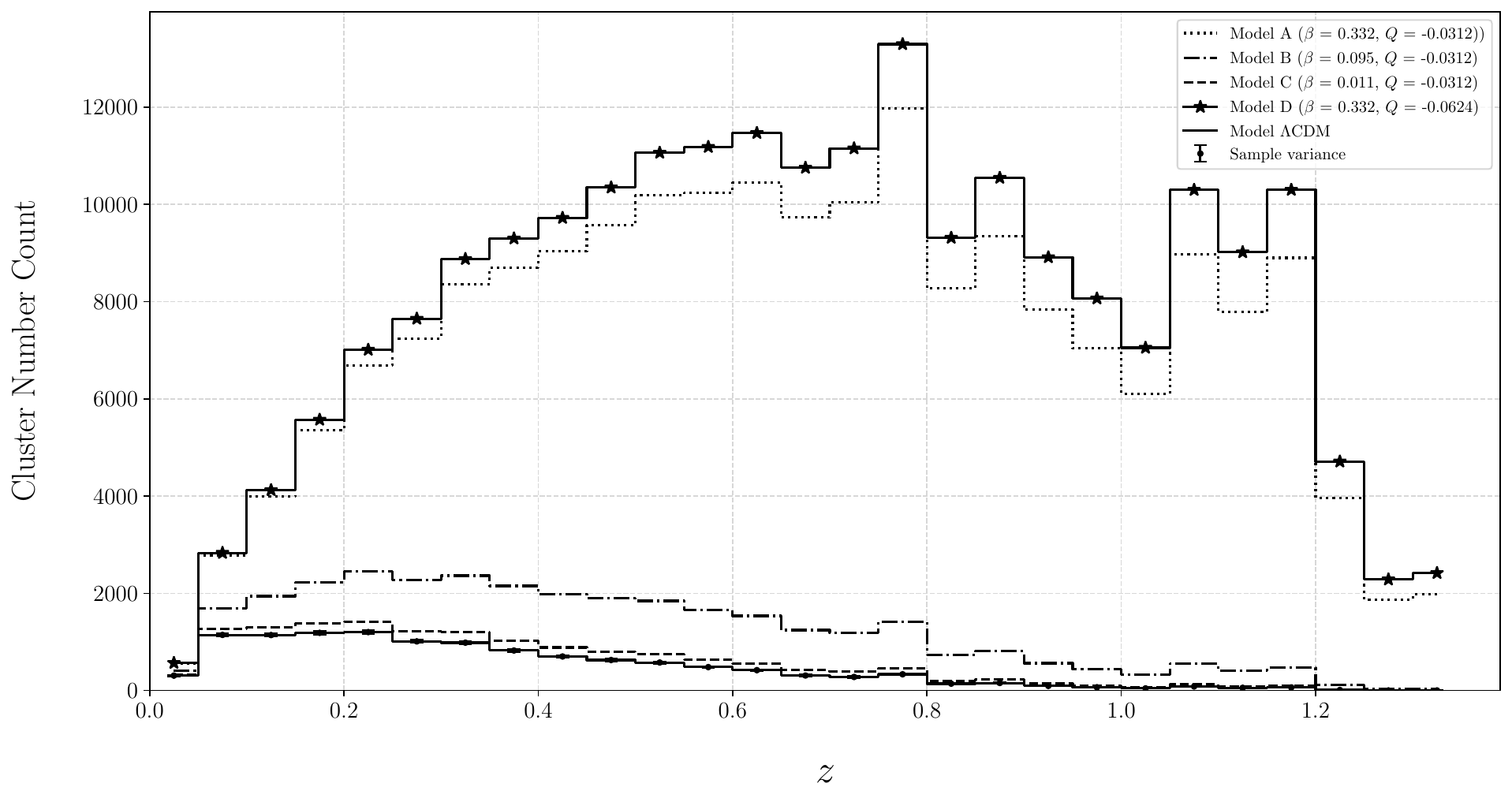}
\caption{Plots of the number of clusters $N$ per redshift bin $\Delta z$, using the PS mass function.
The lines A, B, C, D, and $\Lambda$CDM represent the number of clusters for the models A, B, C, D, and $\Lambda$CDM. 
The errors in the $\Lambda$CDM are estimated from Poisson error in cluster number counts from eROSITA survey data.}
\label{dNdzABC_PS}
\end{figure}

\begin{figure}
\centering
\includegraphics[height=0.5\textwidth, width=0.8\textwidth, angle=0]{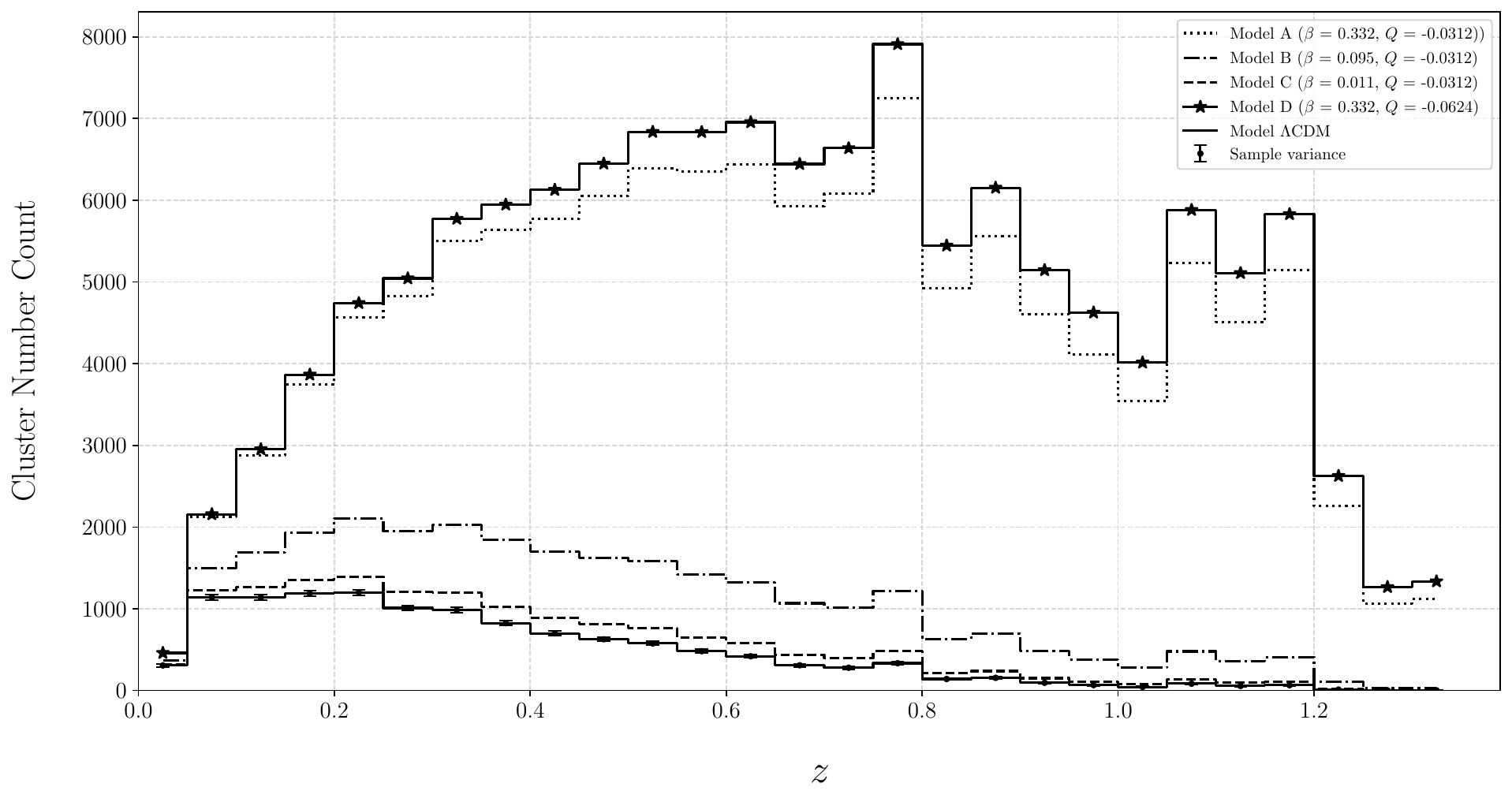}
\caption{Plots of the number of clusters $N$ per redshift bin $\Delta z$, using the ST mass function.
The lines A, B, C, D, and $\Lambda$CDM represent the number of clusters for the models A, B, C, D, and $\Lambda$CDM. 
The errors in the $\Lambda$CDM are estimated from Poisson error in cluster number counts from eROSITA survey data.}
\label{dNdzABC_ST}
\end{figure}

In the calculation of cluster number counts, we set the sky coverage $f_{\rm sky} = 0.31$ $(\approx 12,791$ deg$^2$), according to the eROSITA surveys~\cite{erosita}.
However, the mass limit $M_{\rm min}$ for each redshift bin is not available in the eROSITA surveys data.
In principle, the minimum mass of the clustering regions that can be identified as a cluster is varies across different redshifts.
Hence, $M_{\rm min}$ should not be the same for all redshift bins;
rather, its value is chosen such that the number of clusters computed from Eq.~\eqref{dNdz} for the $\Lambda$CDM model matches the observational results from the eROSITA surveys~\cite{erosita}.
From Figs.~\ref{dNdzABC_PS} and \ref{dNdzABC_ST}, we observe that the number of clusters per redshift bin is substantially enhanced as the momentum coupling $\beta$ or the magnitude of $Q$ increases, compared with the $\Lambda$CDM model.
This feature appears for both the PS and ST mass functions because the enhancement arises from the suppression of the ratio $\delta_{\crit}^2 / \sigma^2(R, z)$ in the exponential term.
Hence, we may conclude that the coupling between dark energy and dark matter leads to an enhancement of cluster number counts independent of the choice of analytic mass function. 
This is because the exponential factor representing the Gaussianity of the matter density field under the assumptions of the Press-Schechter formalism should generally appear in the mass functions. 
Even though the momentum transfer leads to the suppression of the growth of density perturbations on small scales, which potentially alleviates $\sigma_8$ tension, 
it enhances the number of clusters per redshift interval, which may exceed the observational bound from the eROSITA surveys. 
Since a halo mass function specifically calibrated for coupled dark energy models is not yet available, the predicted cluster abundances are subject to modeling uncertainties associated with the halo mass function itself.  In addition, the present analysis does not incorporate the full cosmology dependence of the survey selection function, including the mass-observable relation and mass calibration.  Consequently, the comparison with the eROSITA cluster counts presented there should be regarded as an illustrative assessment of the impact of momentum coupling on cluster abundance, rather than a rigorous cosmological likelihood analysis. Although the difference between the coupled dark energy models and the reference $\Lambda$CDM prediction exceed the Poisson uncertainties estimated from the observed cluster counts, this should not be interpreted as a statistically rigorous inconsistency with the eROSITA data.  A robust statistical comparison would require a halo mass function calibrated for coupled dark energy cosmologies together with a realistic treatment of the survey selection function and observational covariance, which is beyond the scope of the present work.
Although the present work is limited to theoretical predictions, the formalism developed here is readily applicable to future analyses incorporating realistic survey selection functions, weak-lensing mass calibration, and galaxy clustering measurements \citep{Allen_ea2011}. Future spectroscopic surveys such as DESI will provide precise measurements of the growth of cosmic structure through full-shape galaxy clustering and redshift-space distortions \citep{DESI2016}. These observations offer complementary constraints on the growth history and peculiar velocity field, which are directly affected by momentum coupling in the dark sector. In addition, weak-lensing observations from DES \citep{DES2005}, KiDS \citep{KiDS2013}, HSC \citep{HSC2018}, and upcoming and ongoing surveys such as Rubin LSST \citep{LSST2009}, Euclid \citep{Euclid2020}, and Roman \citep{ROMAN2015} will significantly improve cluster mass calibration and the characterization of survey selection functions. These improvements will reduce the dominant systematic uncertainties in cluster abundance analyses and enable more robust tests of momentum-coupled dark energy models. A joint analysis combining galaxy clustering, redshift-space distortions, weak lensing, and cluster abundance therefore represents a promising direction for future work.


\section{Conclusions}
\label{sec5}
We explore the influences of energy and momentum coupling between dark energy and dark matter on structure formation by using the spherical collapse model and the analytic mass function to estimate cluster number counts.
As the momentum coupling parameter $\beta$ increases,
the growth rate of matter perturbations on small scales is suppressed, 
consequently decreasing the value of $\sigma_8$.
We find that the extrapolated linear density contrast at the collapse redshift is also suppressed as the strength of momentum coupling increases. 
Since the extrapolated linear density contrast is suppressed much more than the growth rate of matter perturbation,
the magnitude of the mass function at a particular redshift is enhanced with the increasing strength of momentum coupling. 

Using both the PS and ST mass functions, 
we compute cluster number counts by setting the mass limit for each redshift bin such that the number counts in each bin for the $\Lambda$CDM model match the results from the eROSITA survey data.
We find that the predicted number of clusters per redshift interval increases as the momentum coupling strength increases.
Even when the strength of momentum coupling is set to the lower bound of 2$\sigma$ in \cite{bestfit}, 
the predicted number counts in the coupled dark energy model remain much larger than the Poisson error in the number counts from the eROSITA survey. 
This can be the consequence of uncertainties arising from computing the number counts for the coupled dark energy model using the spherical collapse model and the PS and ST mass functions.


\section*{Acknowledgements}
We would like to thank Nicolas Clerc for his helpful suggestions.  This work was supported by NARIT through the Scholarship for Research student, and was received funding support from the NSRF via the Program Management Unit for Human Resources \& Institutional Development, Research and Innovation [grant number B37G660013]. 

\appendix
\section{Appendix}
\subsection{Dimensionless Variables}
\label{appen1}

In this work, we follow \cite{scaling} to write the background quantities in terms of the following dimensionless variables: 
\ba
x &\equiv& \frac{\dot{\phi}}{\sqrt{6}\,HM_{\rm pl}}\,,\qquad 
y \equiv \frac{M_{\rm pl}\,e^{- \lambda \phi/(2 M_{\rm pl})}}{\sqrt{3}H} \,, 
\label{xydef}\\
\Omega_{\phi} &\equiv& q_s\, x^2 + \tilde{y}^2, \qquad \Omega_m \equiv ( 1+f_1 ) \frac{\rho_m}{3\,M_{\rm pl}^2\, H^2},\,
\label{omegacandb}
\ea
where
\be
q_s \equiv 1 + 2\beta\,, \qquad
\tilde{y} = \frac{\sqrt{V_0}}{M_{\rm pl}^2} y \,.
\ee
The second-order time derivative of the scalar field and the time derivative of the Hubble parameter are
\ba
\label{epsilonbg}
\epsilon_{\phi} &\equiv& \frac{\ddot\phi}{H\dot\phi}
= - 3 + \frac{\sqrt{6}}{2 q_s x}\, (\lambda y^2 - Q\Omega_m) \,,\\
\label{xibg}
\xi &\equiv& \frac{\dot H}{H^2}
= - 3 q_s x^2 - \frac{3}{2} \Omega_m\,.
\ea
The expressions for the energy density and equation of state parameter of the scalar field is
\ba
\rho_{\phi} &=& q_s\, \frac12 \,\dot\phi^2 + V_0\, e^{-\lambda \phi / M_{\rm pl}} = q_s x^2 + \tilde{y}^2 \,,\\
w_{\phi} &=& \frac{q_s\, \dot\phi^2 / 2 - V_0\, e^{-\lambda \phi / M_{\rm pl}}}{q_s\, \dot\phi^2 / 2 + V_0\, e^{-\lambda \phi / M_{\rm pl}}} = \frac{q_s x^2 - \tilde{y}^2}{q_s x^2 + \tilde{y}^2} \,.
\ea 

\subsection{Coefficients of the Evolution Equation}
\label{apx2}

The coefficients of the linear evolution equation for the density contrast are given by \cite{scaling} 
\be
\nu=\frac{2m \beta (1+2\beta)(5+\xi+2\epsilon_{\phi})\,x^2 
+(2+\xi+\sqrt{6}\,Qx)[1+(2-m)\beta]\,\Omega_m}
{2m \beta (1+2\beta) x^2 + \Omega_m 
[1+(2-m)\beta]}\,,
\label{coefnu}
\ee
and
\be
G_{cc} = \frac{1+r_1}{1+r_2} \frac{1}{8\pi M_{\rm pl}^2} \,,
\label{def:gcc}
\ee
with
\ba
r_1 &=& \frac{2Q[3Q \Omega_n+\sqrt{6} m \beta x 
(2+\epsilon_{\phi}+\sqrt{6}Q x)]}
{3\Omega_m [1+(2-m)\beta]}\,,\\
r_2 &=& \frac{2m \beta (1+2\beta)\, x^2}
{\Omega_m [1+(2-m)\beta]}\,.
\ea
\vspace{\baselineskip}
\subsection{Comparison with the Case of Non-Identical Baryons and CDM collapse}
\label{compare}
\begin{figure}
\centering
\includegraphics[height=0.4\textwidth, width=0.6\textwidth, angle=0]{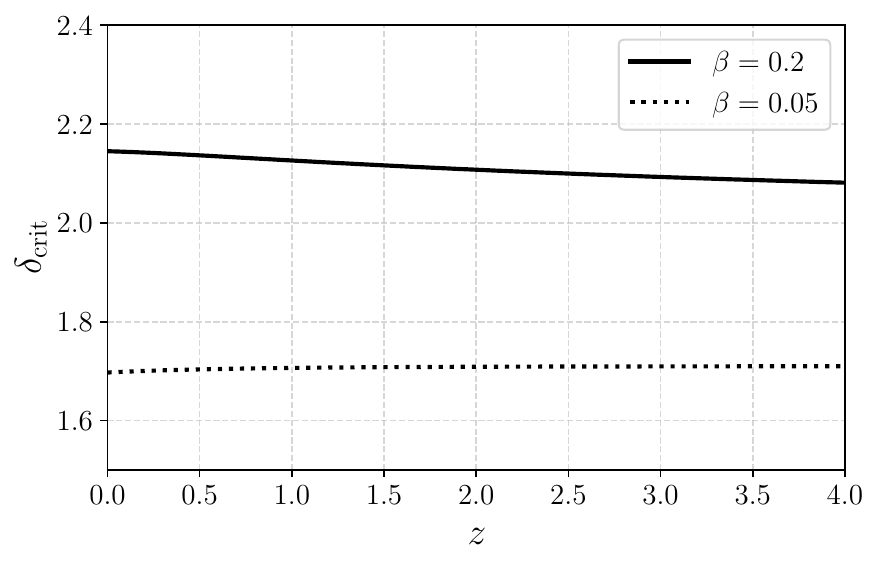} 
\caption{Redshift dependence of the extrapolated linear density contrast $\delta_{\crit}$.}
\label{fig:dc1}
\end{figure}

\begin{figure}
\centering
\includegraphics[height=0.4\textwidth, width=0.6\textwidth, angle=0]{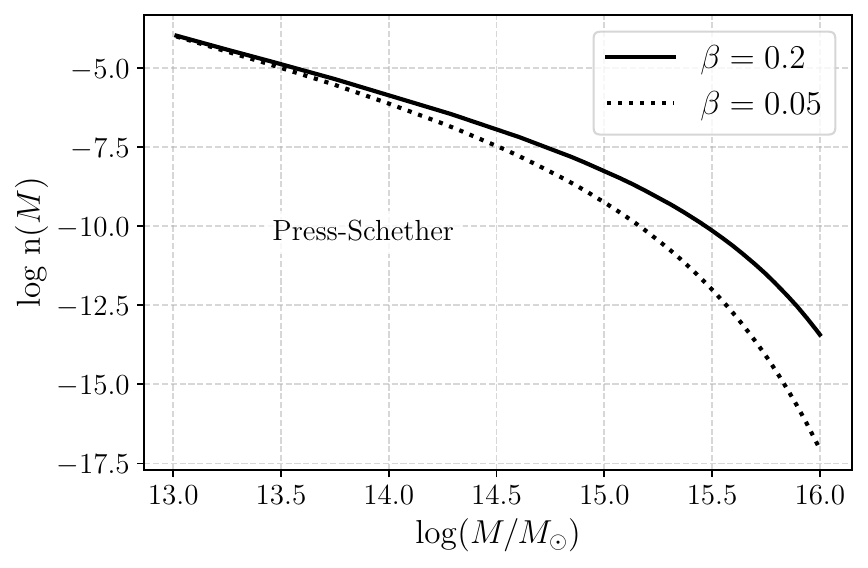}
\caption{Number density as a function of mass  at z = 0, obtained from the PS mass function.}
\label{fig:n(M)_PS}
\end{figure}

\begin{figure}
\centering
\includegraphics[height=0.4\textwidth, width=0.6\textwidth, angle=0]{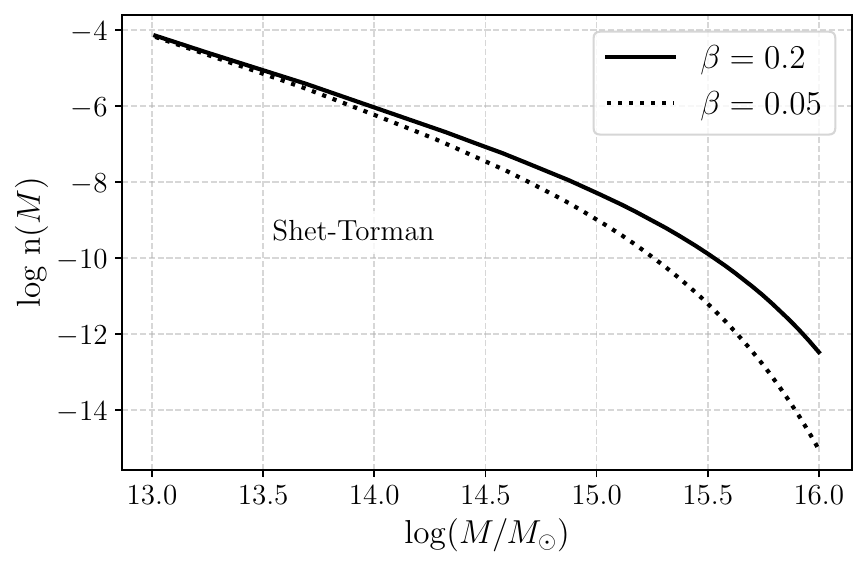}
\caption{Number density as a function of mass at z = 0, obtained from the ST mass function.}
\label{fig:n(M)_ST}
\end{figure}

\indent
We use the evolution equations for the background universe, density perturbation, and spherical collapse from \cite{Mainini:2006zj} to compute extrapolated linear density contrast and number density of clusters based on our simplification used in the main analysis.
In those equations, there are no momentum coupling and the strength of the energy transfer is denoted by $\beta = - Q$, where $Q$ is the strength of energy coupling in our analysis.
In Fig.~\ref{fig:dc1}, the extrapolated linear density contrast for two values of $\beta$ are shown.
For $\beta = 0.2$, $\delta_{\rm crit}$ differs from that in figure 5 of \cite{Mainini:2006zj} by 10 percent,
while the difference is less than  1 percent for $\beta = 0.05$.
The cluster number density at redshift zero for the PS and ST mass functions are shown in Figs.~\ref{fig:n(M)_PS} and~\ref{fig:n(M)_ST}.
The cluster number density from our simplification is not much different from those in the figures 7 and 8 of \cite{Mainini:2006zj}.


\newpage

\end{document}